\documentstyle[12pt]{article}
\include{epsf}
\newcommand{\bc}{\begin{center}}
\newcommand{\ec}{\end{center}}
\newcommand{\bt}{\begin{tabular}}
\newcommand{\et}{\end{tabular}}
\newcommand{\bdes}{\begin{description}}
\newcommand{\edes}{\end{description}}

\newcommand{\be}{\begin{equation}}
\newcommand{\ee}{\end{equation}}
\newcommand{\bea}{\begin{eqnarray}}
\newcommand{\eea}{\end{eqnarray}}
\newcommand{\non}{\nonumber}

\newcommand{\half}{\frac{1}{2}}
\newcommand{\ba}{\begin{array}}
\newcommand{\ea}{\end{array}}

\newcommand{\bsig}{\mbox{\boldmath $ \sigma $}}

\newcommand{\bu} { {\bf u}}

\newcommand{\bn} { {\bf n}}

\newcommand{\btt} { {\bf t}}

\newcommand{\bx} { {\bf x}}
\newcommand{\bee}{ {\bf e}}

\newcommand{\dou}{\partial}
\newcommand{\leftjb} {[\![}
\newcommand{\rightjb} {]\!]}
\newcommand{\ju}[1]{ \leftjb #1 \rightjb }
\newcommand {\eqn}[1] {eq.\/~(\ref{#1})}
\newcommand {\upp} {\bu_P}
\newcommand {\uoo} {\bu_O}
\newcommand {\duu} { \Delta \bu}
\newcommand{\taua}{\tau_{app}}

\renewcommand{\thefigure}{\arabic{page}}
\newcommand{\figtex}{Shenoy and Phillips, Figure \thefigure}

\title{ {\sc Finite Sized Atomistic Simulations \\ of Screw Dislocations} }
\author{ { {\sc Vijay B. Shenoy}  $\;$  {\sc and} $\;$ {\sc Rob Phillips}} \\ {\em Division of Engineering, Brown University, Providence, RI 02912} } 
\date{}

\begin{document}
\baselineskip=24pt
\begin{titlepage}
\begin{figure*}
\centerline{\epsfysize=1.0in \epsfbox{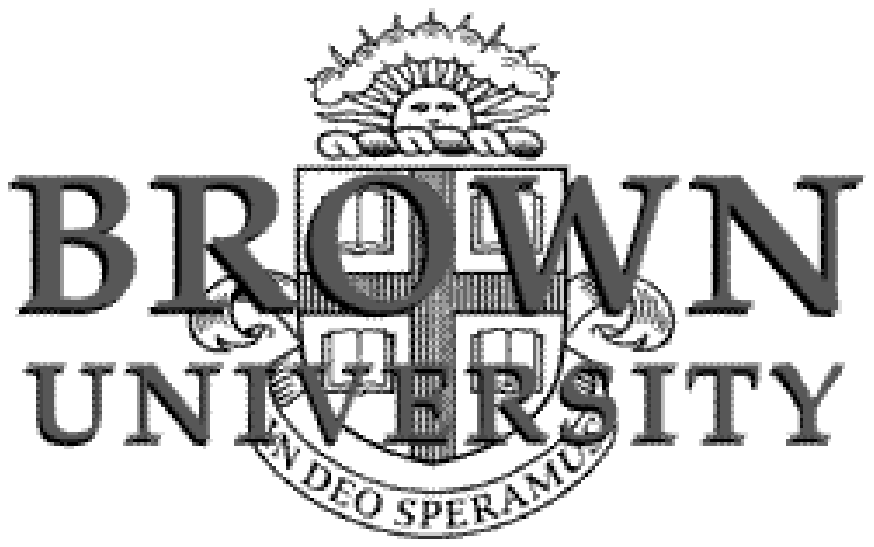}}
\end{figure*}
\Huge
\bc
Finite Sized Atomistic Simulations \\ of Screw Dislocations 
\ec
\Large
\vspace{2.5in}
\bc
{ \sc Vijay B. Shenoy  and  Rob Phillips }
\ec
\vspace{3.0in}
\bc
Atomistics Group \\
Division of Engineering \\
Brown University \\
Providence, RI 02912
\ec
\end{titlepage}
\maketitle

\begin{abstract}
The interaction of screw dislocations with an applied stress is studied using atomistic simulations in conjunction with a continuum treatment of the role played by the far field boundary condition. A finite cell of atoms is used to consider 
the response of  dislocations to an applied stress and this introduces an additional force on the dislocation due to the presence of the boundary. Continuum mechanics is used to calculate the boundary force which is subsequently accounted for in the equilibrium condition for the dislocation. Using this formulation, the lattice resistance curve and the
associated  Peierls stress are  calculated for screw dislocations in several
close packed metals. As a concrete example of the boundary force method, we compute the bow out of a pinned screw dislocation;  the line-tension of the dislocation is calculated from the results of the atomistic simulations using a variational principle that explicitly accounts for the boundary force.
\end{abstract}

\section{Introduction}

Atomistic simulations with carefully chosen potentials have been shown to account for the structure and energetics of a range of extended defects (Vitek 1992). Despite these
successes, one of the main drawbacks of atomistic 
simulations is that they are limited to small simulation cells,
necessitating an approximate treatment of  boundary conditions. A popular method used in the simulation of dislocations is to use a {\em finite sized } simulation cell and to adopt the continuum linear elastic solution for the defect of interest as the initial trial solution for the atomistic energy minimisation with the additional
constraint that the atoms on the boundary of the cell be fixed at the
positions dictated by the elastic solution. The rationale behind this approach is that at large distances from the core of the defect, continuum linear elasticity provides an accurate description of the displacement, strain and stress fields. This method has been applied with much success to the determination of the static core structure of dislocations (Vitek 1992). Upon application
of an applied stress, the defect of interest will adopt a new 
equilibrium configuration.  In particular, the dislocation core
moves relative to the axis of symmetry of the simulation
cell. This motion is influenced by the boundary, i.e., the defect encounters {\em an additional force} that depends on its position relative to the boundary of the cell. 
Failure to account for this boundary effect in the analysis of the interaction of the
defect with an applied stress leads to results that are simulation cell-size
dependent making the {\em quantitative} determination of mechanical parameters such as 
the line tension and the Peierls stress of the dislocation difficult. Indeed, Basinski, Duesbery and Taylor
(1971) observed such effects in their study of the interaction of screw 
dislocations with an applied stress in a sodium lattice. 
Although the importance of the boundary stress has been realised,
a systematic analysis  of its effects has been wanting.

In this paper we develop a method that explicitly accounts for the boundary force 
using linear elasticity, for the case of screw dislocations. It is shown that the boundary force scales as the inverse square of the size of the simulation cell. This formulation is  used in conjunction with atomistic simulations to obtain the lattice resistance function for screw dislocations in several close packed metals.  In addition, the bow out of a pinned screw dislocation under the influence of an applied stress is studied using a variational principle that incorporates the boundary force, demonstrating that in this case the synthesis of atomistic and continuum methods yields a consistent description of bow out.  One outcome of this calculation is that the line tension of the dislocation can be
computed directly from atomistics. In the following section we present the derivation 
of the boundary force after discussing its origin and significance.
Section 3 contains the results of the lattice resistance calculations, 
while the atomistic study of dislocation bow out may be found in 
Section 4. We conclude
with a discussion of the significance of our findings.

\section{Simulation methodology and the boundary force}

Our objective is the simulation of the interaction of an infinite straight screw dislocation,
in an infinite crystal, with an applied stress. To model this
problem using a finite simulation cell we choose a cylindrical cell of atoms with
axis along the $x_2$-direction (cf. fig. 1). The $x_1x_2$ plane is the slip plane,
the $x_2$-direction is the line direction as well as the Burgers vector direction 
of the dislocation. A screw dislocation is placed at the center of this cell at $O$
by displacing the atoms from their perfect lattice sites according to the continuum
linear elastic solution for a Volterra dislocation and this configuration is used as 
the trial solution for the atomistic energy minimisation. The minimisation is performed keeping the atoms in 
region $F$  fixed at the positions given by the linear elastic solution; only the atoms in  
region $M$ ({\em dynamic region}, a cylinder of radius $R$) enclosed by $F$ are allowed to move.
The relaxed core of the dislocation is thus obtained. The interaction of the dislocation with a 
homogeneous applied shear stress is simulated by displacing the atoms in both $F$ and $M$ 
by the displacement fields
corresponding to the homogeneous strain that is equivalent to the stress of interest.
Again, the atomistic energy minimisation step is performed keeping the atoms in the region $F$ fixed 
and allowing the atoms in the region $M$ to move. On performing the above procedure,
the dislocation moves through the crystal to a new position $P$, 
which is determined by the equilibrium condition
\bea
\taua + \tau_b(d) + \tau_L(d) = 0, \label{leq}
\eea
where $d = \overline{OP}$ as shown in fig. 1, $\taua$ is the resolved shear 
stress which effects a Peach-Koehler force $\taua b$ in the $x_1$-direction
($b$ is the magnitude of the Burgers vector), $\tau_L$ is the {\em 
lattice resistance function} (Kocks, Argon and Ashby 1975) which is 
periodic in $d$  and $\tau_b$ is the {\em boundary stress} which is a monotonically increasing (in magnitude)  function of $d$. 

We now explore the origin and significance of the boundary stress term. Upon application of a homogeneous strain, the net displacement of the atoms before 
relaxation in region $F$ and on the external boundary $\dou M_e$ of $M$ is equal to the sum of the elastic displacements due to the dislocation at $O$ ($\bu_O$) and the displacements due to the homogeneous strain as dictated by linear elasticity.  After
relaxation, the dislocation moves to  point $P$; if the dislocation was set
 in an infinite crystal, the displacements of the atoms in region $F$ would have been
 equal to the sum of the elastic displacements due to the presence of the
 dislocation at $P$ ($\bu_P$) and the displacements due to the homogeneous strain. Since the atoms in the region $F$ are held fixed  during the relaxation, the boundary condition {\em after relaxation} is not consistent with the fields of the dislocation at $P$. This inconsistency causes a spurious increase in the total energy stored in the system, i.e., there is an additional energy cost in moving the dislocation from $O$ to $P$. This additional energy produces an energetic stress  on the dislocation (in the sense of Eshelby 1951) which we call the {\em boundary stress}.  This boundary stress 
 depends on the radius $R$ of the dynamic region, making the quantities calculated from the results of the simulations, cell-size dependent. The boundary stress is particularly important in the simulations of dislocations in crystals with low lattice resistance such
 as those with the fcc and hcp structures.  In materials with high lattice resistance such
 as bcc metals this term may be neglected as the lattice resistance term will be 
 several orders of magnitude larger than the boundary term.

In order to derive the boundary stress, we assume that the total energy $\cal E$ is built up of an elastic energy $E$ and a misfit energy $\cal M$, i.e.,
${\cal E} = E + {\cal M}$. The misfit energy accounts for the energy required to maintain the slip distribution due to the presence of the dislocation, and the elastic energy is due to the elastic straining of 
the crystal and may be thought of as arising from a continuous distribution
of infinitesimal dislocations.
Thus the net change in the energy ($\Delta \cal E$) on moving the dislocation to the point $P$ is then given by $\Delta E(d) + \Delta {\cal M}(d)$.
In the absence of the inconsistent boundary condition, the change in elastic energy would vanish ( in accordance with a Peierls-Nabarro type model which states that the elastic energy remains unchanged as the dislocation moves through the crystal). We assume that the presence of the inconsistent boundary does not effect  the misfit energy, and thus the additional energy due to the inconsistent boundary is entirely elastic. In addition, we assume that the change in elastic energy $\Delta E$ computed using a {\em Volterra model} of the dislocation gives a good approximation to this energy change.  These assumptions are valid when the size of the simulation cell ($R$) is large compared to the core width of the dislocation. Under these assumptions, the change in elastic energy $\Delta E$ may be computed and the boundary stress is given by
\bea
\tau_b(d) = - \frac{1}{b}\frac{\dou \Delta E}{ \dou d }, \label {bfeq}
\eea 
and the lattice resistance function $\tau_L$ may  be obtained as
\bea
\tau_L(d) = - \frac{1}{b}\frac{\dou \Delta {\cal M} }{ \dou d }. \label {lrfeq}
\eea
In the following discussion,
the external boundary of the dynamic region is denoted by $\dou M_e$, the slip surface
$\overline{AO}$ is denoted by $\dou M_s$
and $\dou M_{s'}$ denotes the slip surface $\overline{AP}$ that corresponds 
to the dislocation at $P$.
 All the linear elastic fields associated with the presence of the dislocation
 at $O$ are denoted by a subscript $O$ (e.g. $\bu_O$) while a subscript $P$ is used
 to denote fields  associated with the dislocation  at $P$. In both
 cases, the fields correspond to an unconstrained dislocation
 in an infinite crystal.  The dislocation is displaced from $O$ to $P$, while
 keeping the displacements on $\dou M_e$ fixed at $\bu_O$ (a {\em constrained} motion of the dislocation). This corresponds to the process that occurs in the atomistic simulation when the dislocation moves under the  influence of the applied stress.
It is clear that $\Delta E$ is the difference in the elastic energy $E_2$ of the system
when the dislocation is at $P$ after the constrained motion and the elastic energy $E_1$ when the dislocation is at $O$. The energy $E_1$ is evaluated as
\bea
 E_1 = \half\int_{\dou M_s}\btt_O \cdot \ju{\bu_O} dS + \half\int_{\dou M_e} \btt_O \cdot \bu_O dS,                  \label{en1} 
\eea
where $\btt_O$ is the traction vector on the  relevant boundary due to the stress field of the dislocation at $O$ ( $\btt_O = \bsig_O \cdot {\bf n}$ where $\bsig_O$ is the stress tensor and ${\bf n}$ is the normal to the boundary/surface),  and $\ju{\;}$ represents the jump in the displacement across the slip surface (i.e., the
 Burgers vector).  
The total energy $E_2$ of the configuration with the dislocation at $P$ (after the constrained motion) may be evaluated as follows. This configuration is taken as the superposition of the  fields $\upp$ due to the presence of the dislocation at $P$ as though in an infinite crystal, and $\duu$ which is an equilibrium displacement field with 
\bea
\duu = \uoo - \upp, \; \; \; \; \forall \bx \in  \dou M_e \label{dbc}
\eea
and with the condition that  $\ju{\duu} = {\bf 0}$ on $\dou M_{s'}$.
It must be noted that $\duu$ at a point in the {\em interior} of $M$ is {\em not} equal to $\uoo - \upp$. The energy in the displaced configuration is
\bea
 E_2 = \half\int_{\dou M_{s'}} \btt_P \cdot \ju{\bu_P} dS + \half\int_{\dou M_e} \btt_P \cdot \bu_P dS + \int_{\dou M_e} \btt_P \cdot \Delta \bu dS +\half \int_{\dou M_e} \Delta \btt \cdot \Delta \bu dS,                \label{en2} 
\eea 
where $\btt_P$ is the traction associated with the stress field of the dislocation at $P$ (as though in an infinite body), $\Delta \btt$ is the traction associated  with the fields $\duu$. The additional energy $\Delta E$ stored in the system due to the boundary condition on $\dou  M_e$ may be expressed as
\bea
\Delta E & = & E_2 - E_1 = \Delta E^a + \Delta E^b + \Delta E^c, \label{tote} 
\eea 
with the three terms being given by
\bea
\Delta E^a & = & \half\int_{\dou M_{s'}} \btt_P \cdot \ju{\upp} dS - \half\int_{\dou M_{s}} \btt_O \cdot \ju{\uoo} dS, \label{wa} \\
\non \\
\Delta E^b &=& \int_{\dou M_e} \btt_P \cdot (\uoo  - \half \upp ) dS \label{wb} - \half\int_{\dou M_e} \btt_O \cdot \bu_O dS, \\
\non \\
\Delta E^c &=& \half \int_{\dou M_e} \Delta \btt \cdot \duu dS  .\label{wc}
\eea
We choose to partition the energy change in this fashion for convenience as it will be shown below how these three contributions may each be evaluated separately.

In preparation for  evaluating the energy difference, we record the  isotropic, linear elastic
displacement and stress fields for a Volterra screw dislocation (Hirth and Lothe 1968)  at a point $Q$ with coordinates $(x_Q, 0)$
\bea
(u_2)_Q(x_1, x_3)  = u_Q(x_1, x_3) = -\frac{b}{2 \pi} \arctan{ \left( \frac{x_3}{x_1 - x_Q}\right) }.
\eea
The point $Q$ is a generic point on the slip plane; $Q$ may, for example, be taken to be $O$ or $P$.
The components of the stress tensor $\bsig_Q$ are given by   
\bea 
(\sigma_{12})_Q (x_1, x_3) = \frac{\mu b}{ 2 \pi}  \left( \frac{\sin{\phi}}{r'}\right),  \;\;\;\;\;\;\;  (\sigma_{32})_Q (x_1, x_3) =  -\frac{ \mu b}{ 2 \pi} \left( \frac{\cos{\phi}}{r'}\right)     \label{s32}
\eea
where $\phi = \arctan(x_3/(x_1 - x_Q))$,  $r' = \sqrt{(x_1-x_Q)^2+x_3^2}$ and $\mu$ is the shear modulus. All other displacement and stress components not symmetric to those given above vanish.  In the developments that follow, we freely switch between cartesian $(x_1,x_3)$ and polar coordinates $(r,\psi)$. 
The notation is simplified by adopting the convention that  
\bea
\sigma_Q=\btt_Q \cdot {\bf e}_2 = (\bsig_Q \cdot {\bf n}) \cdot {\bf e}_2 = (\sigma_{12})_Q n_1 +  (\sigma_{32})_Q n_3   \label{sigq},
\eea
where $\bf n$ is the normal to the pertinent boundary and $\bee_i$ denotes the basis vector along the $i^{\mbox{th}}$ coordinate axis.

First we compute $\Delta E^a$ which requires the evaluation of surface integrals over the domains $\dou M_{s}$ and $\dou M_{s'}$ for which $\bn = {\bf e}_3$. Taking $r_o$ to be the elastic core cut-off radius, we see that
\bea
\half\int_{\dou M_{s'}} \btt_P \cdot \ju{\upp} dS = \frac{ \mu b^2}{4 \pi} \ln{\frac{R + d}{r_o}}, \;\; \mbox{and} \;\; \half\int_{\dou M_{s}} \btt_O \cdot \ju{\uoo} dS =   \frac{ \mu b^2}{4 \pi}  \ln{\frac{R}{r_o}}.  \non 
\eea
 It is now immediate that the
contribution of these terms to the energy difference is
\bea
\Delta E^a =  \frac{ \mu b^2}{4 \pi}\ln(1 + a), \label{wa1} \label{fwa}
\eea
with $ a = d/R $.

To evaluate $\Delta E^b$ we note that the unit normal to the external boundary $\dou M_e$ is given by $\bn( \psi ) =  \cos{\psi} {\bf e}_1 + \sin{\psi} {\bf e}_3$. 
Using eqs. \ref{s32} and \ref{sigq} it follows that $\btt_O \cdot {\bf e}_2 = \sigma_O = 0$ on $\dou M_e$
and hence from \eqn{wb}
\bea
\Delta E^b = \int_{\dou M_e} \btt_P \cdot (\uoo  - \half \upp ) dS  = \int_{\dou M_e} \sigma_P ( u_O - \frac{1}{2} u_P) dS. \label{iwb}
\eea
Since in this case  $x_P = d$, one obtains that
\bea
\sigma_P(R, \psi) = \frac{\mu b d}{2 \pi} \frac{ \sin{\psi}}{ {r'}^2}.
\eea
for points with coordinates $(R,\psi)$ on $\dou M_e$ noting that
$\phi = \arctan{(\sin{\psi}/(\cos{\psi} - a)) }$ and
$r' = \sqrt{R^2 + d^2 - 2 R d \cos{\psi}}$.
Substituting all relevant quantities into \eqn{iwb}, results in 
\bea
\Delta E^b = -\frac{\mu b^2 }{4 \pi^2} a I_1(a)  \label{fwb}
\eea
where
\bea
I_1(a) = \int_{-\pi}^{\pi} \left[ \left(  \psi - \half \arctan{ \left( \frac{\sin{\psi}}{\cos{\psi} -a } \right) }\right) \frac{\sin{\psi}}{{\rho'}^2} \right]  d \psi, 
\eea
and $ {\rho'}^2 = ( 1 + a^2 - 2 a \cos{\psi})$.

We now turn to the  evaluation of  $\Delta E^c$. From \eqn{wc} we have that
\bea
\Delta E^c = \half \int_{\dou M_e} \Delta \btt \cdot \duu dS = \half \int_{\dou M_e} \Delta \sigma \Delta u dS \label{iwc}
\eea
with $\Delta \bu = \Delta u {\bf e}_2$ and $(\Delta \bsig \cdot {\bf n})\cdot {\bf n} = \Delta \sigma $. 
The  evaluation of  $\Delta E^c$, requires both knowledge of $\Delta u$ on $\dou M_e$, which is provided by boundary conditions, and of $\Delta \sigma$.  Since  $\Delta u$ is an equilibrium field, it satisfies the elastic equilibrium equation (Navier's equation) in $M$,  
\bea
\nabla^2 ( \Delta u) &  =  & 0
\eea
with the associated boundary condition that on $\dou M_e$ 
\bea
\Delta u & = & - \frac{b}{2 \pi}\left( \arctan{ \left( \frac{x_3}{x_1} \right)} - \arctan{ \left( \frac{x_3}{x_1 - d} \right)} \right). \label{sbc} 
\eea
This Dirichlet problem for $\Delta u$ in the interior of $M$ may be solved using the Poisson integral formula (John 1982), i.e.,
\bea
\Delta u(r, \psi) = -\frac{b}{4 \pi^2} \int_{-\pi}^{\pi} \left[\left( \theta - \arctan{ \left( \frac{\sin{\theta}}{\cos{\theta} -a }\right)} \right)\left( \frac{ R^2 - r^2}{R^2 + r^2 - 2rR \cos{(\psi  - \theta)}}\right) \right] d \theta \label{dus}
\eea
where $(r, \psi)$ are the polar coordinates of an interior point of $M$. To evaluate $\Delta \sigma$ on $\dou M_e$ we evaluate the strain field corresponding to the displacement field given by \eqn{dus} and use the elastic constitutive equations. This results in
\bea
\Delta \sigma & = & \frac{\mu b}{2 \pi^2 R} J(\psi) \label{dsig} 
\eea
with $J(\psi)$ defined by
\bea
J(\psi)  & = & \lim_{\rho \rightarrow 1} \frac{2 \pi^2}{b}\frac{ \dou \Delta u( \rho, \psi)}{\dou \rho} \non \\
&= & \lim_{\rho \rightarrow 1} \int_{-\pi}^{\pi} \left[\left( \theta - \arctan{ \left( \frac{\sin{\theta}}{\cos{\theta} -a }\right)} \right)\left( \frac{2 \rho - (\rho^2 + 1) \cos{(\psi - \theta)}}{(1 + \rho^2 - 2\rho \cos{(\psi  - \theta)})^2}\right) \right] d \theta,
\eea
where $\rho$ is the normalised value of $r$, i.e., $\rho = r/R$.
Substituting eqs. \ref{sbc} and \ref{dsig}  into \eqn{iwc} we find
\bea
\Delta E^c = - \frac{\mu b^2}{8 \pi^3} I_2(a), \label{fwc}
\eea
with
\bea
I_2(a) = \int_{-\pi}^{\pi} J(\psi) \left( \psi - \arctan{ \left( \frac{\sin{\psi}}{\cos{\psi} -a } \right)} \right) d \psi .
\eea

Having evaluated the various terms that contribute to the elastic energy 
as a result of the inconsistent boundary condition, 
we are now in a position to evaluate the boundary stress.
Collecting terms $\Delta E^a$, $\Delta E^b$ and $\Delta E^c$ from
eqs. \ref{fwa}, \ref{fwb} and \ref{fwc} one obtains that
\bea
\Delta E  & =  &\frac{\mu b^2}{4 \pi^2} Z(a) 
\eea
where $Z(a)$ is given by
\bea
Z(a)  & = & \pi \ln{(1+a)} - aI_1(a) - \frac{1}{2 \pi} I_2(a). 
\eea
It is remarkable that $\Delta E$ does not depend explicitly on $d$ and $R$, but depends
only on their ratio $a$. Also, the result is independent of the
elastic core cutoff radius $r_o$ (on the assumption that it remains fixed). 

Although the above discussion was restricted to isotropic linear elastic
solids, our method may be applied to the anisotropic case as well.
We indicate the changes that are required to accomplish this in the 
case of hcp crystals with $\langle 11\overline{2}0 \rangle$ type screw 
dislocations. In this case the simulation cell is chosen to be an
elliptic cylinder, i.e., the boundary $\dou M_e$ satisfies the equation
\bea
x_1^2 + \frac{x_3^2}{k^2} = R^2,
\eea
where $k = \sqrt{ C_{44}/C_{66}}$ and $C_{ij}$ have their usual meanings (Hirth and Lothe 1968). The terms $\Delta E^a$ and $\Delta E^b$ can be evaluated  straightforwardly by following
the steps of the previous section. To evaluate $\Delta E^c$ we note that $\Delta u$ satisfies
\bea
\frac{\dou^2 \Delta u}{\dou x_1^2} + k^2 \frac{\dou^2 \Delta u}{\dou x_3^2} = 0
\eea
with boundary conditions as in eqs. \ref{dbc}. This equation may be solved after a coordinate transformation $(x_1,x_3) \rightarrow (y_1, y_3)$ where $y_1 = x_1$ and $y_3 = x_3/k$; $\Delta u$ is harmonic in the new coordinates and $\dou M_e$ transforms to a circle of radius $R$, and thus $\Delta u$ may be calculated using the Poisson integral formula. The traction $\Delta \sigma$ is calculated using the elastic constitutive equations and $\Delta E^c$ is evaluated as before. The final expression for $\Delta E$ may be written in this case as
\bea
\Delta E  & =  &\frac{C_{44} b^2}{4 \pi^2} Z(a,k)  \label{anise}
\eea
where in this case the dimensionless
energy change $Z$ depends on both $a$ and the ratio $k$ which is a measure of the elastic
anisotropy. The case of isotropic solids is obtained from \eqn{anise} by using $k = 1.0$ and $C_{44} = \mu$.

The boundary stress may now be obtained using \eqn{bfeq} as
\bea
\tau_b(a) =  -\frac{C_{44} b}{4 \pi^2 R} \; \frac{\dou Z(a, k)}{\dou a} =  -\frac{C_{44} b}{4 \pi^2 R} \; F(a,k) \label{bfeq1}, 
\eea
where $F(a,k)$ is the derivative of $Z$ with respect to $a$. The dimensionless stress $F$ is plotted in fig. 2. For small values of $a$, $Z$ admits a quadratic approximation of the form
\bea
Z(a,k) = \half A(k) a^2, \label{quada}
\eea
which demonstrates that  when $a \ll 1$ (i.e., $d \ll R$) the boundary stress is linear in the dimensionless parameter $a$. This linear nature of $F$ in the indicated regime is evident from fig. 2. The constant $A$, which depends on the value of $k$, is  the {\em boundary stress coefficient} and the boundary stress $\tau_b$ becomes
\bea
 \tau_b(d) =  -\frac{C_{44} b A(k)}{4 \pi^2 R^2} d \label{bforc}
\eea
The values of the boundary stress coefficient are computed to be 6.28 for the isotropic case and 6.19 for the case of titanium (HCP). It is clear from \eqn{bforc} that the  boundary force is directly proportional to $d$ for a given $R$  and for a given $d$ depends on the inverse square of the size of the simulation region $M$. Assuming that a typical Burgers vector is of the order of 5.0\AA $\;$ and that the radius of the simulation cell is about 50.0\AA $\;$ one obtains a boundary stress of about 0.00025$\mu$, when $d$ is taken to be one Burgers vector. Thus the boundary effects become highly significant in the cases of 
materials with low lattice resistance (fcc and hcp metals).
On the other hand, in the case of BCC screw dislocations, the boundary term may be neglected in comparison with the Peierls stress which is two orders of magnitude larger.

In light of our development of the
boundary stress, we now proceed to derive the equilibrium condition (\eqn{leq}). We
assume that the dislocation has moved to the point $P$ under the influence of the applied stress $\taua$ and is in equilibrium. The principle of virtual work states that if the dislocation is given an arbitrary virtual displacement $\delta$ from $P$, the net external work done will
 equal the change in the net energy stored (per unit thickness along the $x_2$-direction), i.e.,  
\bea
\underbrace{\taua b \delta}_{\mbox{external work}} = \underbrace{\frac{ \dou \Delta E}{\dou d} \delta + \frac{ \dou \Delta {\cal M}}{\dou d} \delta }_{\mbox{change in energy}} . \label{che}
\eea
Eq. \ref{che} can be rewritten as
\bea
(\taua + \tau_b(d) + \tau_L(d)) \delta = 0. 
\eea
Making use of eqs. \ref{bfeq} and \ref{lrfeq}, and noting that
$\delta$ is arbitrary, the equilibrium condition (\eqn{leq}) is obtained. 
By virtue of the continuum arguments outlined here, we are now in
a position to use the boundary force in the context of explicit
atomistic simulation.

It may be noted that, in the isotropic case, the boundary force can derived by
placing an image dislocation at a point $(R^2/d,0)$ on the $x_1$ axis. The 
resulting fields obtained by superposition will then satisfy the condition
\eqn{dbc} and the boundary force is now equal to the force exerted by the image
dislocation on the dislocation at $P$. The advantage in using the energetic
method of obtaining the boundary force is the generalization to the anisotropic
case, and possibly for other defects such as the edge dislocation.

\section{Lattice resistance functions and Peierls stress}

In this section we present the results of the calculation of the lattice resistance function for screw dislocations in several close packed metals. As mentioned above, the form of the lattice resistance function was discussed by Kocks, Argon and Ashby (1975). Although
atomistic calculations have been used previously
to compute the Peierls stress which corresponds to
a single point on the lattice resistance curve, it is
one of the aims of the present work to compute  the entire 
lattice resistance  function  from
atomistic simulation. This is achieved using the following method. The position of the dislocation ($d$) is monitored as a function of the applied stress. The applied stress is sufficiently
small so that the displacement $d$   is small compared with  $R$; thus the linear boundary stress approximation may be applied (\eqn{bforc}). From the equilibrium condition  (\eqn{leq}) the lattice resistance function can be obtained as
\bea
\tau_L(d) = -(\taua + \tau_b(d)). \label{latr}
\eea 

The motion of the dislocation is monitored relative to an initial
equilibrium position labeled as point $O$ which is chosen to correspond to a Peierls valley. 
To obtain the position $d$ of the dislocation the following procedure is adopted.
The slip distribution on the slip plane $(x_1x_2)$ i.e., distribution of $\ju{u_2}$ as a function of $x_1$ is obtained using the method given by Miller and Phillips (1996). The position $d$ of the dislocation is taken to be that value of $x_1$ at which $\ju{u_2}$ equals one half the Burgers vector ($b/2$).  For this procedure to be meaningful it is essential that as the dislocation moves through the crystal, the slip distribution moves in an essentially self similar fashion, which we find is a good approximation in the cases considered here.

To obtain the relationship between the applied stress $\taua$ and $d$, the stress is  applied in small increments and the configuration is relaxed at every step as explained before. Once the applied stress reaches a desired maximum value (this is chosen so that the resulting
$d/R$ value  is not overly large), the stress is decremented until it reaches zero, i.e., the direction of loading is reversed. Fig. 3 shows the net resistance that the dislocation would have to overcome in its journey through the crystal (i.e., $\tau_b + \tau_L$). The expected
$\taua$ vs. $d$ curve is $OABCDEFG$ as shown in fig. 3 in the loading branch and $GHIJKLMO$ during the unloading process. Thus if the applied stress is sufficiently large, the dislocation travels through various Peierls wells (for example $OA$ and $BC$ are in different Peierls wells, as are $IJ$ and $KL$ in the unloading branch). It is clear that the $\taua$ vs. $d$ curve obtained  will depend on the simulation cell size.  On the other hand, if the  boundary term is subtracted off using \eqn{latr}, the results should be found to be cell size independent. Also the lattice resistance function computed from various Peierls wells must agree (since the lattice resistance is periodic), in addition to being simulation cell size independent.

To examine the validity of our ideas concerning the boundary force, we have examined the response of screw dislocations to an applied stress in aluminum and titanium. The point of these calculations is to revisit the idea of lattice resistance of dislocations from an atomistic perspective while approximately accounting for the finite simulation cell size.

Our first test case was the simulation of a $\frac{a}{2} \langle 110 \rangle$ type screw dislocation in fcc aluminum which was performed using the embedded atom potentials developed by Ercolessi and Adams (1993). For details on the embedded atom method, see Daw and Baskes (1984). Fig. 4  shows the relationship between $\taua$ and $d$ (compare with fig. 3). It is clear that this curve is simulation cell-size dependent. The lattice resistance function computed using \eqn{latr} is given in fig. 5. In these calculations the value of the boundary force coefficient was taken to  be 6.3. It is seen that the computed lattice resistance function is simulation cell-size independent, with a resulting Peierls stress of $0.00068 \mu$.

Our additional test was made on hcp titanium. In this case, embedded atom potentials developed by Igarashi, Khantha and Vitek (1991) were used. The screw dislocations simulated had a Burgers vector of $\frac{a_o}{3} \langle 11 \bar{2}0 \rangle$ on the basal plane.  The lattice resistance function is plotted in fig. 6.
 The anisotropy  ratio $k$ for Titanium is $1.06$ and the boundary force coefficient
 $A$ for this value of $k$ is  6.2. This value for $A$ was found to be too high; and when $A$ was set to 5.5 the lattice resistance obtained from different Peierls wells turned out to be consistent. The value of the Peierls stress is about $0.00034 \mu$.

It must be noted that in plotting the lattice resistance functions in figs. 5 and 6 using \eqn{latr}, the negative sign appearing in the equation has been dropped. Thus the resistance function must be interpreted as the external stress necessary to overcome the lattice resistance. In both the cases discussed above, the lattice resistance functions from different Peierls wells are plotted in the same graph. It is thus evident that the computed lattice resistance functions are periodic. It is interesting to note that in all the cases, the lattice resistance functions are multi-welled.
The screw dislocations in all the materials discussed above undergo a splitting reaction into Shockley partials and thus the lattice resistance curves are for these extended dislocations. The effect of this splitting on the boundary force is discussed in Appendix 1.  The value of $A$ based on isotropic elasticity works well for the case of Al while in the case of Ti the value of $A$ had to be set to 5.5 (as opposed to 6.2 given by the analytic calculation) to obtain consistent values from different Peierls wells (i.e., for the resistance function to be periodic). 

\section{Bow out of a pinned dislocation segment}
The simulation of the bow out of a pinned dislocation segment is an important step towards  understanding the mechanics of  Frank-Read sources. Work in this direction has hitherto been restricted to continuum formulations. Foreman (1967) applied Brown's self-stress (Brown 1964) approach to simulate dislocation bow out. Assuming that the dislocation behaves like a flexible string with a line tension $T$, a simpler theory is obtained (Nabarro 1967, de Wit and Koehler 1959). The latter theory has the advantage of simplicity and  we adopt it for the interpretation of the results of the atomistic simulations.  The goals of these simulation are i) to examine the validity of the concept of line tension from an atomistic viewpoint and to compute its value from ``first principles''
 and ii) to compare the shape of the dislocation bow out predicted by the continuum theory with that obtained from atomistics.

The atomistic simulation of bow out of a pinned segment under an applied stress is achieved as follows. A straight screw dislocation is placed along the axis of a cylindrical cell of atoms of length $L$ (the axis of the cylinder is the $x_2$-axis in fig. 1), with the radius of the dynamic region $M$ set to $R$.  A strain corresponding to the applied stress $\taua$ is imposed on
this configuration and the energy is minimised with respect to the atomic coordinates in the dynamic region. In the energy minimisation step, in addition to the atoms in region $F$, the atoms on the $x_2 = 0$ and $x_2 = L$ planes
are held fixed imitating the pinning effect on the dislocation.

To obtain the equilibrium shape of the bow out as predicted by continuum theory, we consider a screw dislocation on the $x_2$ axis pinned at $x_2 = 0$ and $x_2 = L$, i.e., the length of the pinned segment is $L$. Let the equilibrium shape of the dislocation be given by the function $x_1 = f(x_2$) (cf. fig. 7). The equilibrium shape is given by the function that minimises the functional
\bea
I(f) =  \int_0^L \left( T \sqrt{1 + f'^2} - \taua b f + \frac{\lambda}{2} f^2 \right)  dx_2,  \label{varp}
\eea
 $f'$ is the derivative of $f$ with respect to $x_2$ and $\lambda = A\mu b^2/(4 \pi^2 R^2)$, where $A$ is the boundary force coefficient. The quadratic term  accounts for
 the additional energy due to the fixed boundary and is absent from a standard continuum formulation.  It is assumed in this formulation that the applied stress is larger that the Peierls stress so that the lattice resistance (misfit energy) can be neglected.  In addition, the applied stress $\taua$ is also assumed to be small compared to the shear modulus $\mu$ of the material. The function $f$ satisfies the Euler-Lagrange equation of the variational principle \eqn{varp}, i.e.,
\bea
  \frac{d}{dx_2} \left( \frac{f'} {\sqrt{1 + {f'}^2 }} \right)  + ( \frac{\taua b}{T} -  \frac{\lambda}{T}  f) = 0, \; \; \; \; f(0) = f(L) = 0, 
\eea
the solution of which is (assuming $ | f' | \ll  1$ since $\taua \ll \mu$)
\bea
f(x_2) & = & \frac{\taua b}{\lambda} \left(1 - \frac{ \sinh{\sqrt{ \frac{\lambda}{T}}(L - x_2)} +  \sinh{\sqrt{\frac{\lambda}{T}}x_2}}{ \sinh{\sqrt{\frac{\lambda}{T}}L}} \right)  \label{caten}
\eea
which expresses the fact that under the influence of the boundary stress the shape of the bow out is given by a catenary. 
Noting that the maximum bow out $d$ occurs at $x=L/2$, we find that
\bea
d = f(L/2) = \frac{\taua b}{\lambda}\left[1 -  \mbox{sech}\left(\frac{L}{2}\sqrt{\frac{\lambda}{T}}\right) \right] \label{mbow}.
\eea

The following strategy is adopted in computing the line tension for the results of the atomistic simulations. The maximum bow out $d$ is ``measured'' from the results of an atomistic simulation. If $d$ is known, the line-tension $T$ can be obtained from \eqn{mbow} as
\bea
T = \frac{ \lambda L^2 }{4 \left(\mbox{sech}^{-1}\left[{ \left(1 - \frac{\lambda d}{\taua b} \right) \ } \right] \right)^2 } \label{linet}.
\eea
 If  the effect of the boundary is not explicitly accounted for $(A \rightarrow 0)$, then $d$ is related to $T$ using
\bea
d = \lim_{\lambda \rightarrow 0} \frac{\taua b}{\lambda}\left[1 -  \mbox{sech}\left(\frac{L}{2}\sqrt{\frac{\lambda}{T}}\right)\right] = \frac{\taua b L^2}{8T} \label{cort}
\eea
and one may  obtain an {\em uncorrected} estimate of the line tension as 
\bea
T_u = \frac{\taua b L^2}{8d} . \label{uncor}
\eea

These results were used to perform atomistic simulations of  dislocation bow out for $\frac{a}{2} \langle 110 \rangle$ type screw dislocation in Al using the EAM potentials developed by Ercolessi and Adams (1993). The values of the line tension computed for various lengths of the dislocation at various levels of 
applied stresses are tabulated in tables 1, 2 and 3 (in all these cases the value of $R$ is 40.0\AA).  The values of line tension computed using \eqn{linet},  which incorporate the boundary stress, are length independent. Also, the values do not depend strongly on the values of the applied stress. Table 4 shows the computed value of the line tension using a larger simulation cell ($R =$ 57.0\AA). It is seen that the computed values agree well with those computed using the smaller cell.  In addition, it is seen that the values of line tension computed from \eqn{linet} and those computed from \eqn{uncor} agree when the cell size becomes larger and when the length of the dislocation is made smaller, as is evident from tables 1 and 4. At larger lengths of the dislocation the bow out $d$ becomes larger (for a given value of $\taua$) and the quadratic approximation \eqn{quada} for the additional energy is an underestimate and hence the boundary force is underestimated. Consequently, the line tension is overestimated, as seen in table 3.  A second interesting trend is seen in the estimates for the line tension in the case where the boundary contribution is {\em not} accounted for. Here as the dislocation length increases, so too does the apparent line tension. This effect may be attributed to the boundary force which mimics the effects of a higher line tension preventing the development of a full bow out.

It is also useful to compare the shape of the bow out as predicted by the continuum formulation (\eqn{caten})  and that obtained from atomistics. It is seen that the shape predicted by the continuum theory which accounts for the boundary force is in close agreement with that obtained from atomistics (cf. fig. 8). Thus the continuum formulation (\eqn{varp}) provides a satisfactory model for the study of dislocation bow out. In addition the concept of line tension is found to be meaningful from an atomistic perspective allowing for a determination of the line tension  from ``first principles''. The value of line tension is found to be about 0.42$\mu b^2$ which is close to the conventional approximation of 0.5$\mu b^2$ (Nabarro 1967). It must be noted that this dislocation undergoes a splitting reaction into Shockley partials as described before and thus the computed line tension is for the extended dislocation.  It is noted that even with the boundary force correction, the computed values of the line tension are slightly different for various levels of applied stress. This is believed 
to arise from the neglect of the lattice resistance term.

Fig. 9 shows a plot of $\ju{u_2}$ as a function of the position on the slip plane extracted from the results of an atomistic calculation using the technique given in Miller and Phillips (1996). The value of $d$ is found by sectioning this surface at a value of one half of the Burgers vector. The positions of atoms on the atomic planes that sandwich the slip plane is are plotted in fig. 10. A careful study of this figure brings out the discrete nature of the dislocation bow out and the core structure of a curved dislocation.

\section{Conclusion}

In this paper, we have described a method for accounting for the additional boundary stresses that act on a dislocation due to the use of a finite sized cell in an atomistic simulation. An explicit formula for the boundary stress is derived for the case of the screw dislocation. The formula was derived based on the Peierls-Nabarro model and a linear elastic reckoning of the elastic energy due to an inconsistent boundary.  The boundary stress is proportional to the inverse square of the size of the simulation cell. This formulation was then used to obtain the lattice resistance function for screw dislocations in several
close packed metals. In addition, a simple 3D configuration for the bow out of a screw dislocation was also studied and the line tension of the dislocation was computed for the case of aluminum. 

The explicit accounting of the boundary stress is essential to obtain meaningful {\em quantitative} values from the results of an atomistic simulation. This is evident from the simulation of the  dislocation bow out that is presented here. 
The effects of the far-field boundary conditions may be overcome with the use of large simulation cells when empirical or semi-empirical potentials are used in the simulation, however it proves to be a serious difficulty in calculations using density functional theory. Thus this method might
be especially  useful in those simulations where there are severe restrictions on the size of the simulation cell.

\section*{Appendix : Effect of dislocation splitting on the boundary force}

It was noted in section 3 that the screw dislocations in aluminum and titanium undergo a splitting reaction. Here we compute the boundary force when the dislocation is split assuming   a splitting distance of $2s$ ( $s=4.0$\AA$\;$ for aluminum and $s=6.1$\AA$\;$ for titanium). It is also assumed that this distance does not change as the dislocation moves under the influence of the applied stress - a premise that appears to be  in good agreement with observations.   Under these assumptions one can compute the boundary stress to be
\bea
 \tau_b(d) =  -\frac{C_{44} b A(k,c)}{4 \pi^2 R^2} d \label{bsplf}
\eea
for $d \ll R$, where $c = s/R$. This relation can be derived using equations eqs. \ref{wa}, \ref{wb} and \ref{wc} where $\bu_P$ is taken to be the sum of the displacement fields due to two screw dislocations with Burgers vector $b/2$ located at $x_1 = d-s$ and $x_1 = d+s$. In the derivation the edge part of the Burgers vector of the Shockley partials is neglected, and thus \eqn{bsplf} is approximate. It is seen that the boundary stress coefficient depends on the splitting ratio $c$ in addition to the anisotropy ratio $k$. A plot of the boundary force coefficient as a function of the splitting ratio is given in fig. 11. The value of $A$ is seen to be unaffected when the splitting is small ($c < 0.1$). This condition is satisfied in the  cases considered here and the effect of splitting is thus small.

\section*{Acknowledgements}
We are grateful to V. Bulatov, A. Carlsson, R. Clifton, M. Fivel
and M. Khantha for their valuable comments. Thanks are also due to R. Miller and E. Tadmor for many useful discussions. Support for this work by NSF under Grant No. CMS-9414648 is gratefully acknowledged. We thank the referee for suggesting the image 
dislocation argument to arrive at the result for the boundary stress.

\newpage
\section*{References}
\noindent
\begin{description}
\item{} Basinski, Z. S., Duesbery, M. S., Taylor, R. (1971): {\em Can. J. Phys.}, {\bf 49}, p. 2160. 
\item{} Brown, L. M. (1964): {\em Phil. Mag.}, {\bf 10}, p. 441.

\item{} Daw, M. S.,  Baskes, M. I. (1983): {\em Phys. Rev. Lett.}, {\bf 50}, p. 1285.
\item{} de Wit, G., Koehler, J. S. (1959): {\em Phys. Rev.}, {\bf 116}, p. 1113.
\item{} Ercolessi, F., Adams, J. (1993): {\em Europhys. Lett.}, {\bf 26}, p. 583.
\item{} Eshelby, J. D. (1951): {\em Solid State Physics}, Seitz, F. and Turnbull, D. ed., {\bf 3}, p. 79.
\item{} Foreman, A. J. E. (1967):  {\em Phil. Mag.}, {\bf 15}, p. 1011.
\item{} Hirth, J. P., Lothe, J. (1968): {\em Theory of Dislocations}, McGraw-Hill.
\item{} Igarashi, M., Khantha, M., Vitek, V. (1991): {\em Phil. Mag. B.}, {\bf 63}, p. 603.
\item{} John, F. (1982): {\em Partial Differential Equations}, Springer-Verlag.
\item{} Kocks, U. F., Argon, A. S., Ashby, M. F. (1975): {\em Progress in Material Science}, {\bf 19}, p. 1.
\item{} Miller, R., Phillips, R. (1996): {\em Phil. Mag. A}, {\bf 73}, p. 803.
\item{} Nabarro, F. R. N. (1967): {\em Theory of Crystal Dislocations}, Oxford University Press.
\item{} Vitek, V. (1992): {\em Progress in Material Science}, {\bf 36}, p. 1.
\end{description}

\newpage
\section*{Figure Captions}
\noindent
\begin{description}
\item{Figure 1:} Geometry of the simulation cell illustrating the implementation of a far field 
 boundary condition by freezing atoms in positions dictated by linear elasticity theory.

\item{Figure 2:} Dimensionless stress $F(a,k)$ as a function of dimensionless displacement of dislocation line, shown for different values of the anisotropy ratios.

\item{Figure 3:} Schematic of $\taua$ vs. $d$ curve. ($d_o$ is the repeat lattice distance)

\item{Figure 4:} $\taua$ vs. $d$ as obtained from atomistic simulations for Al (compare with fig. 3) showing size dependent results in the absence of boundary correction.

\item{Figure 5:} Lattice resistance function for screw dislocation in Al accounting for boundary stress and showing system size independence.
 Dashed lines are provided as a schematic indication of the parts of the resistance curve that are inaccessible using this method.

\item{Figure 6:} Lattice resistance function for screw dislocation in Ti.  Dashed lines are provided as a schematic indication of the parts of the resistance curve that are inaccessible using this method.

\item{Figure 7:} Geometry of dislocation bow out.

\item{Figure 8:} Comparison of atomistic and continuum bow out in Al.  $R=40.0$\AA, $L=211.2$\AA, $\taua=0.0025 \mu$.

\item{Figure 9:} Plot of $\ju{u_2}$ on the slip plane. $R=40.0$\AA, $L=211.2$\AA, $\taua=0.01 \mu$.

\item{Figure 10:} Positions of atoms on atomic planes that sandwich the slip plane. $R=40.0$\AA, $L=211.2$\AA, $\taua=0.01 \mu$.

\item{Figure 11:} Variation of the boundary stress coefficient $A$ with normalised splitting distance $c$.

\end{description}

\newpage

\begin{table}
\bc
\bt{|c|c|c|c|}
\hline
$\taua / \mu$ & $d$ (\AA) & $T_u/\mu b^2$ (without boundary effects) & $ T/\mu b^2$ (with boundary effects)  \\
\hline
0.0025 & 0.83 & 0.43 & 0.40 \\
0.0050 & 1.59 & 0.45 & 0.42 \\
0.0100 & 3.40 & 0.42 & 0.39 \\
\hline
\et
\ec
\caption{Bow out and line tension for screw dislocation
in Al. ($L = 57.2$\AA, $R=$40.0\AA) }
\end{table}

\begin{table}
\bc
\bt{|c|c|c|c|}
\hline
$\taua / \mu$ & $d$ (\AA) & $T_u/\mu b^2$ (without boundary effects) & $ T/\mu b^2$ (with boundary effects)  \\
\hline
0.0025 & 2.43 & 0.53 & 0.41 \\
0.0050 & 4.66 & 0.55 & 0.43 \\
0.0100 & 9.11 & 0.56 & 0.44 \\
\hline
\et
\ec
\caption{Bow out and line tension for screw dislocation
in Al. ($L = 108.5$\AA, $R=$40.0\AA) }
\end{table}

\begin{table}
\bc
\bt{|c|c|c|c|}
\hline
$\taua / \mu$ & $d$ (\AA) & $T_u/\mu b^2$ (without boundary effects) & $ T/\mu b^2$ (with boundary effects)  \\
\hline
0.0025 & 5.19  & 0.94 & 0.46 \\
0.0050 & 10.24 & 0.95 & 0.48 \\
0.0100 & 18.74 & 1.04 & 0.57 \\
\hline
\et
\ec
\caption{Bow out and Line Tension for Screw Dislocation
in Al. ($L = 211.2$\AA, $R=$40.0\AA) }
\end{table}

\begin{table}
\bc
\bt{|c|c|c|c|}
\hline
$\taua / \mu$ & $d$ (\AA) & $T_u/\mu b^2$ (without boundary effects) & $ T/\mu b^2$ (with boundary effects)  \\
\hline
0.0025 & 0.88 & 0.42 & 0.40 \\
0.0050 & 1.62 & 0.46 & 0.44 \\
0.0100 & 3.47 & 0.43 & 0.42 \\
\hline
\et
\ec
\caption{Bow out and line tension for screw dislocation in Al.
($L = 58.1$\AA, $R=$57.0\AA) }
\end{table}

\clearpage

\newpage
\setcounter{page}{1}
\pagestyle{empty}

\begin{figure}[h]
\centerline{\epsfbox[35 203 531 682]{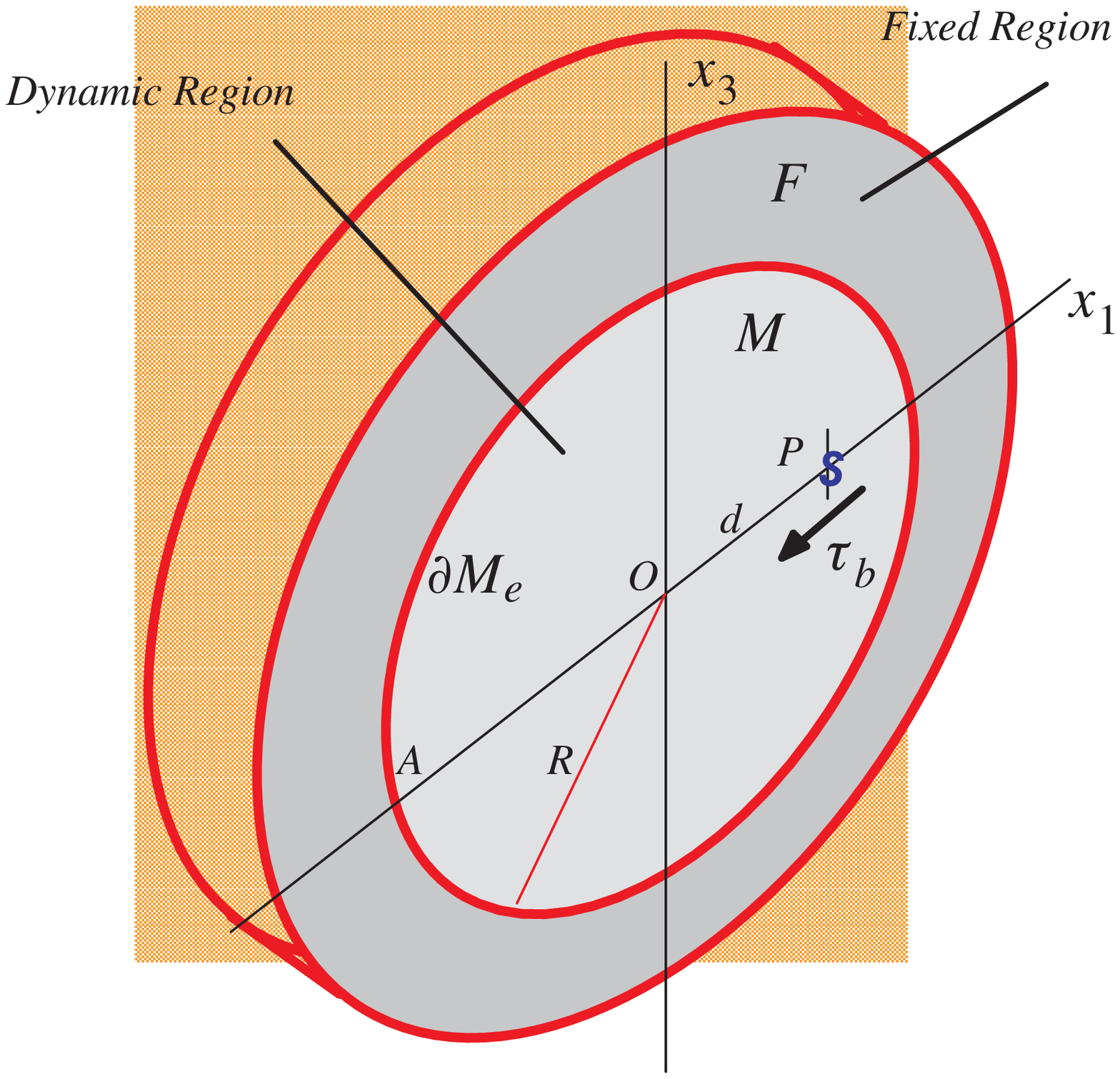}}
\vspace{1.0in}
\centerline{\figtex}
\end{figure}
\clearpage

\newpage

\begin{figure}[h]
 \centerline{\epsfysize=4.0in \epsfbox[44 222 522 573]{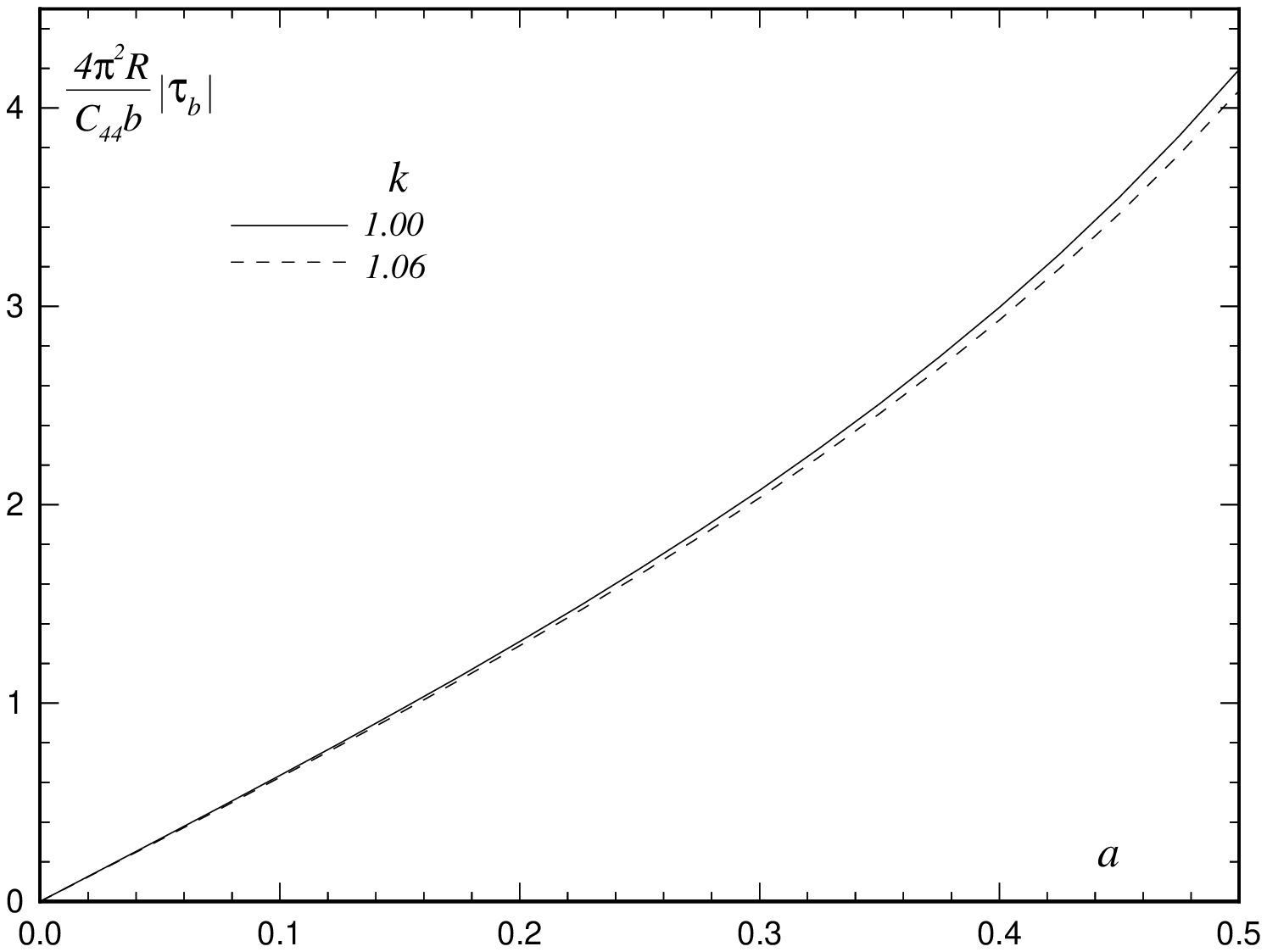}}
\vspace{1.0in}
\centerline{ \figtex}
\end{figure}
\clearpage

\newpage 
\begin{figure}[h]
\centerline{\epsfysize=4.0in \epsfbox[44 222 522 573]{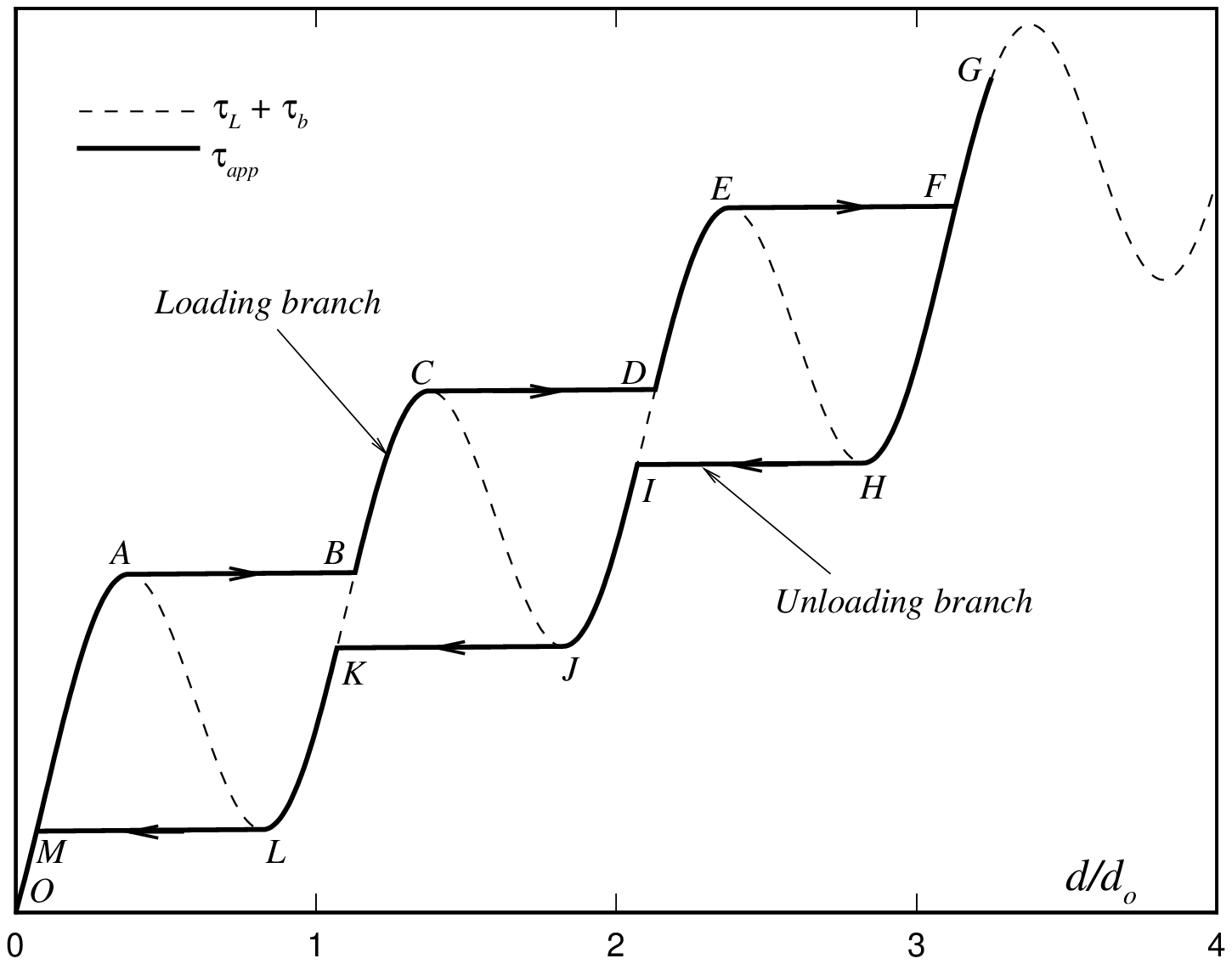}}
\vspace{1.0in}
\centerline{ \figtex}
\end{figure}
\clearpage

\newpage

\begin{figure}[h]
\centerline{\epsfysize=4.0in \epsfbox[44 222 522 573]{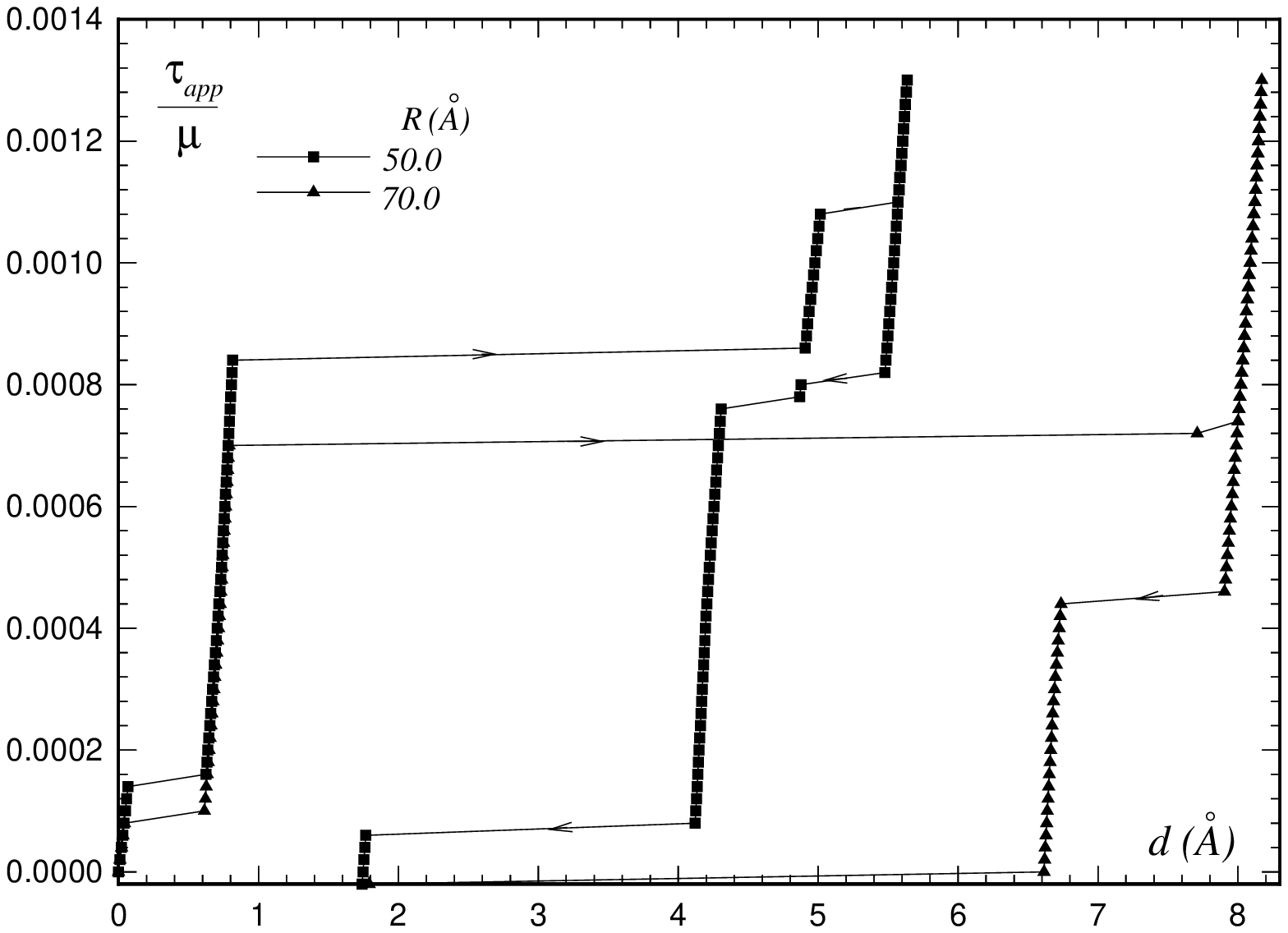}}
\vspace{1.0in}
\centerline{ \figtex}
\end{figure}
\clearpage

\newpage
\begin{figure}[h]
\centerline{\epsfysize=4.0in \epsfbox[44 222 522 573]{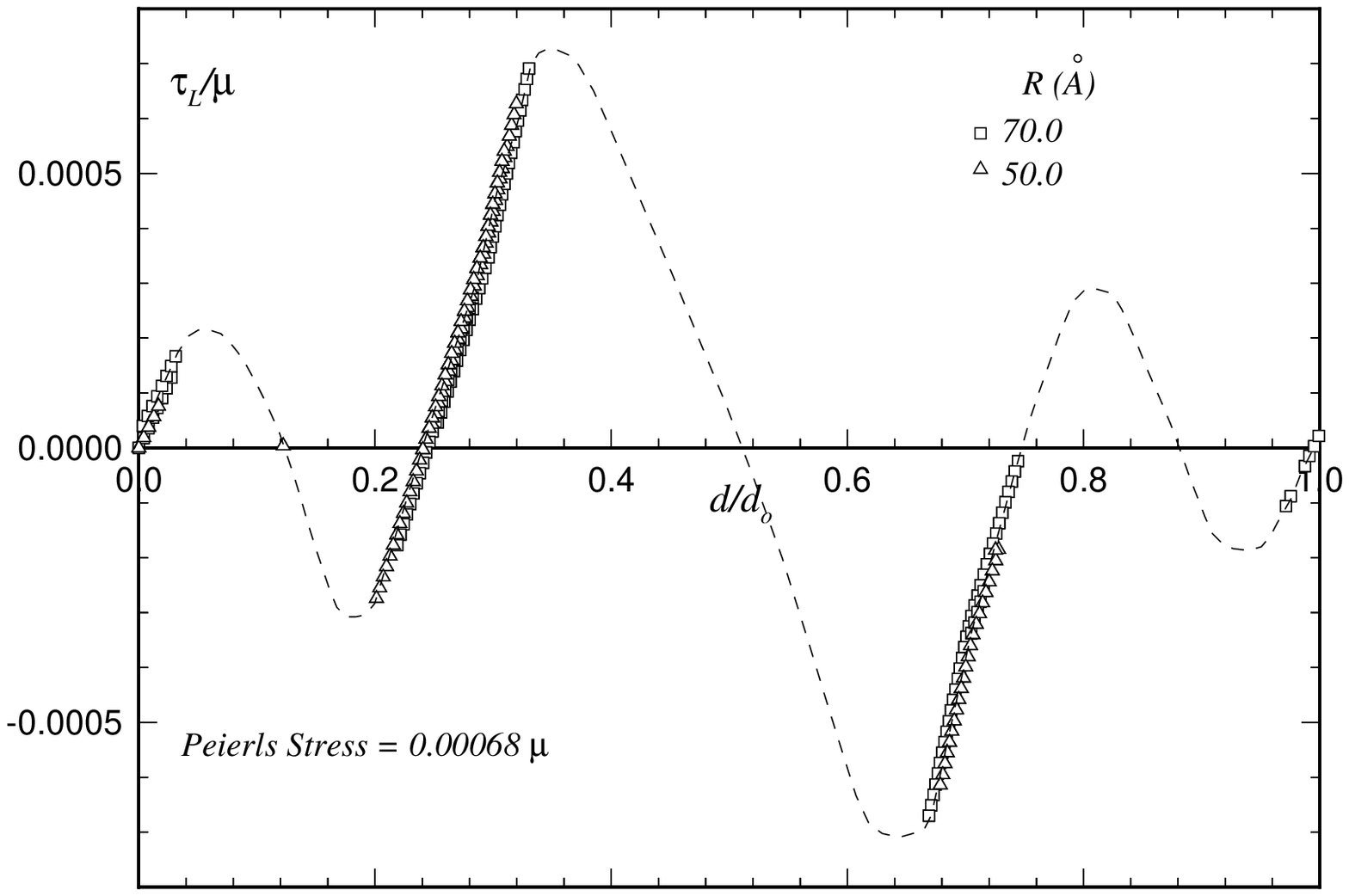}}
\vspace{1.0in}
\centerline{ \figtex}
\end{figure}
\clearpage

\newpage

\begin{figure}[h]
\centerline{\epsfysize=4.0in \epsfbox[44 222 522 573]{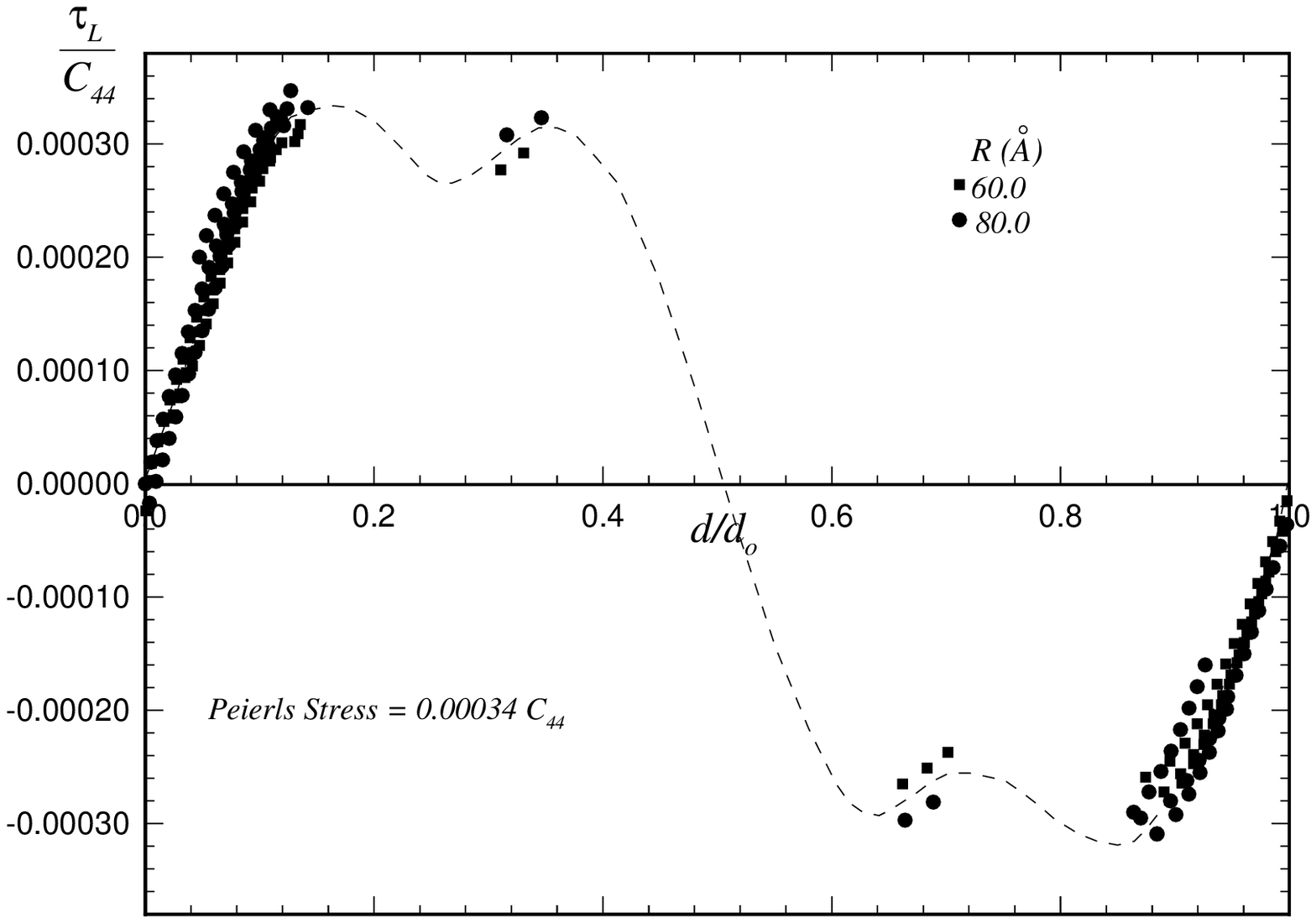}}
\vspace{1.0in}
\centerline{ \figtex}
\end{figure}
\clearpage

\newpage

\begin{figure}[h]
\centerline{\epsfysize=4.0in \epsfbox[44 222 522 573]{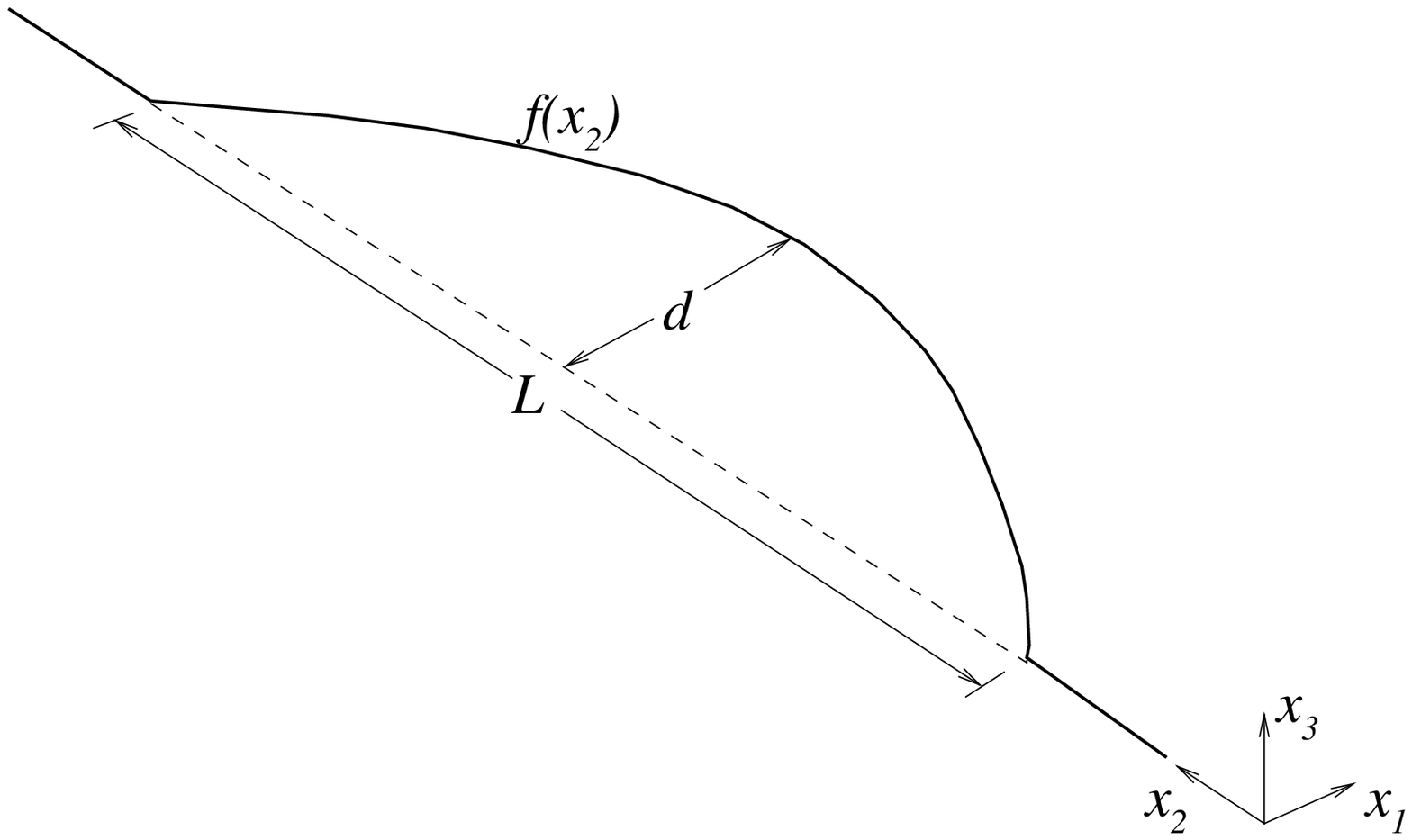}}
\vspace{1.0in}
\centerline{ \figtex}
\end{figure}
\clearpage

\newpage

\begin{figure}[h]
\centerline{\epsfysize=4.0in \epsfbox[44 222 522 573]{bfig13.eps}}
\vspace{1.0in}
\centerline{ \figtex}
\end{figure}
\clearpage

\newpage
\begin{figure}[h]
\centerline{\epsfysize=4.0in \epsfbox{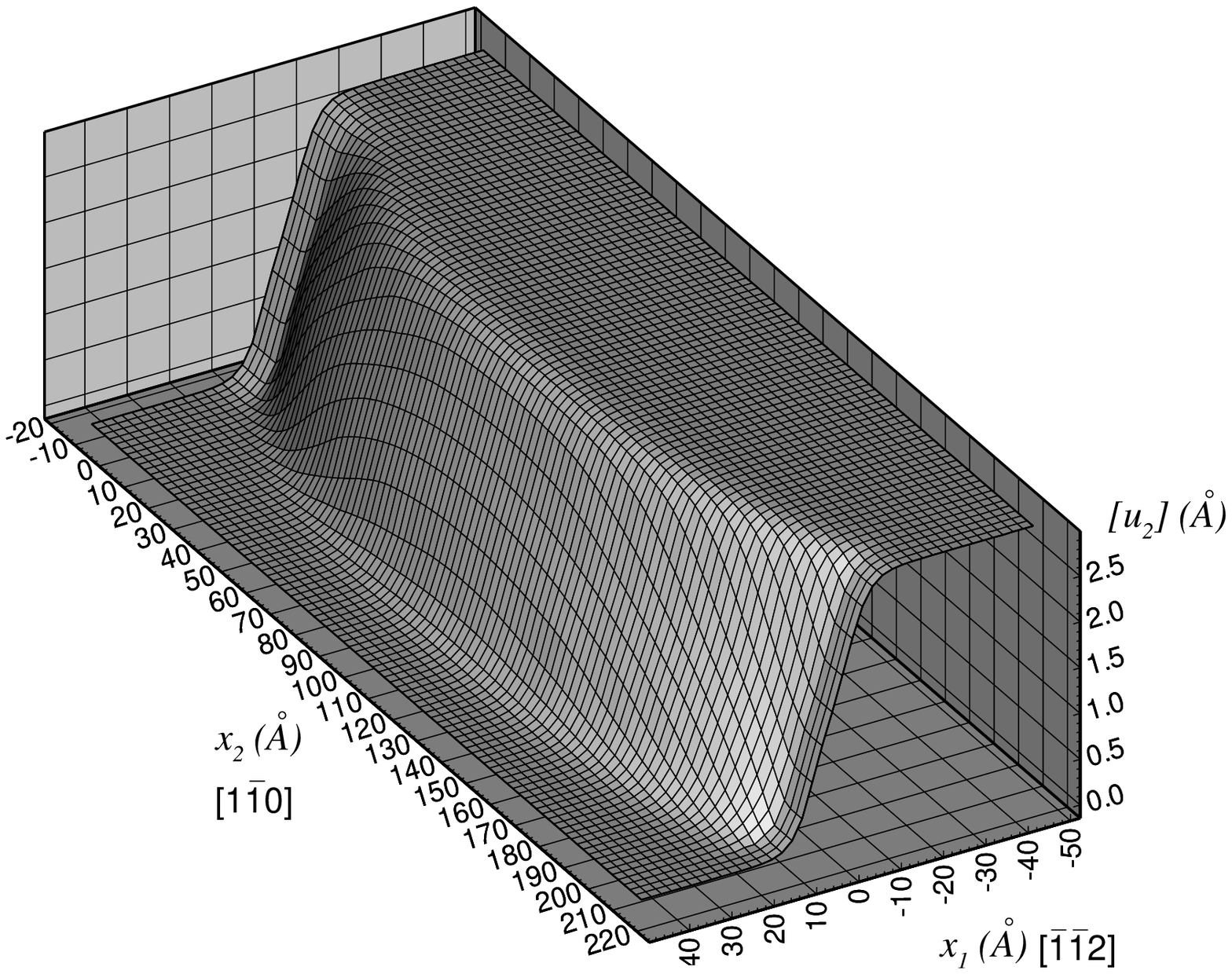}}
\vspace{1.0in}
\centerline{ \figtex}
\end{figure}
\clearpage

\newpage
\begin{figure}[h]
\centerline{\epsfysize=4.0in \epsfbox[32 193 575 517]{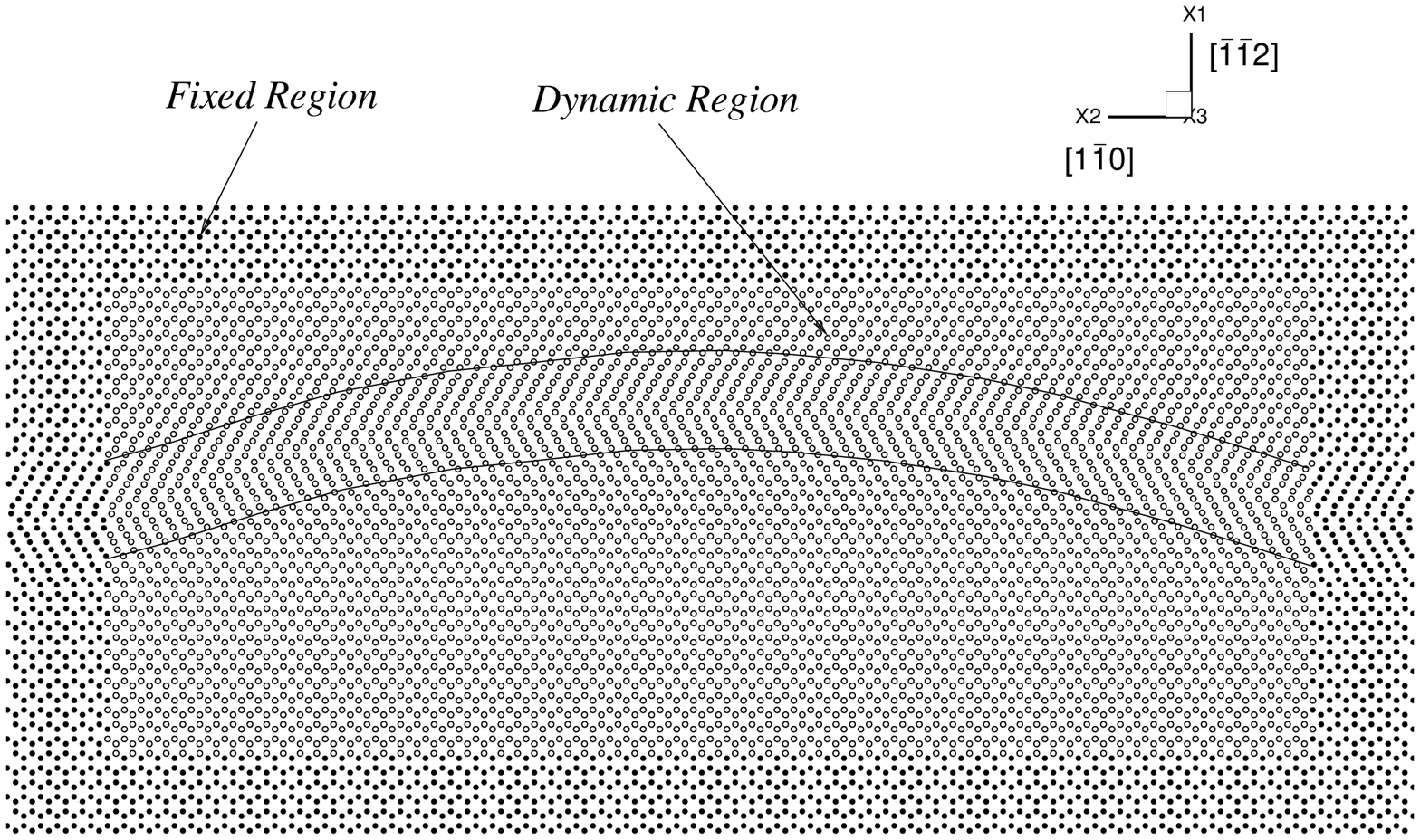}}
\vspace{1.0in}
\centerline{ \figtex}
\end{figure}  
\clearpage

\newpage

\begin{figure}[h]
\centerline{\epsfysize=4.0in \epsfbox[44 222 522 573]{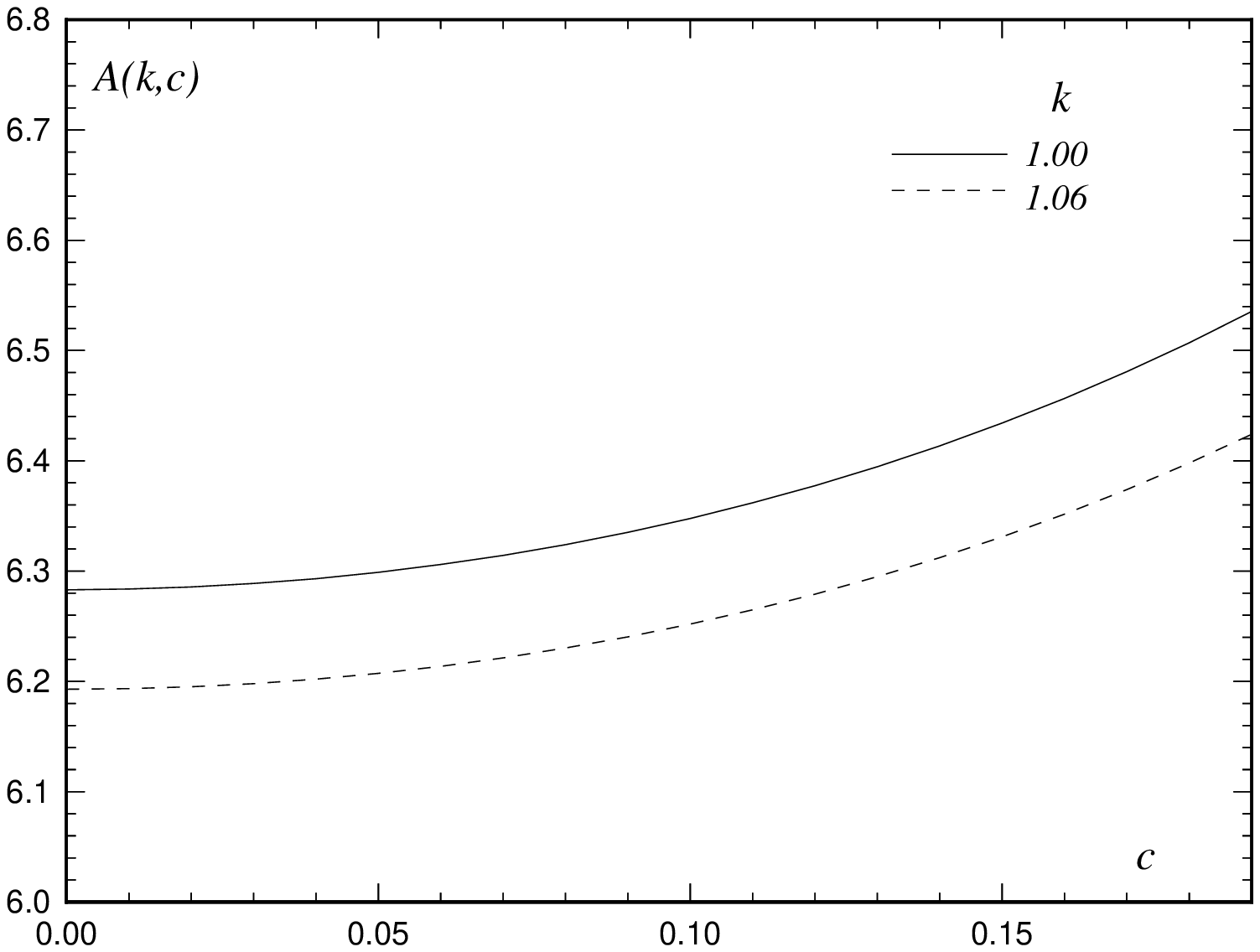}}
\vspace{1.0in}
\centerline{ \figtex}
\end{figure}
\clearpage

\end{document}